\documentclass[3p,twocolumn,longtitle,reversenotenum,sort&compress,lefttitle]{elsarticle}

\pdfoutput=1

\usepackage[utf8]{inputenc}
\usepackage[T1]{fontenc}

\PassOptionsToPackage{hyphens}{url}\usepackage[hidelinks]{hyperref}

\usepackage{libertine}
\usepackage{microtype}
\usepackage{booktabs}

\usepackage{array}
\newcolumntype{L}[1]{>{\raggedright\let\newline\\\arraybackslash\hspace{0pt}}p{#1}}

\makeatletter
\let\@afterindenttrue\@afterindentfalse
\makeatother

\nopreprintlinetrue

\begin{document}

\begin{frontmatter}

\title{Joining Forces for Pathology Diagnostics with AI Assistance:\newline The EMPAIA Initiative}

\author[aff1]{Norman Zerbe}
\author[aff2]{Lars~Ole~Schwen}
\author[aff3]{Christian~Geißler}
\author[aff4]{Katja~Wiesemann}
\author[aff1]{Tom~Bisson}
\author[aff5]{Peter~Boor}
\author[aff1]{Rita~Carvalho}
\author[aff1]{Michael~Franz}
\author[aff1]{Christoph~Jansen\corref{corresp}}
\ead{christoph.jansen@charite.de}
\author[aff1]{Tim-Rasmus~Kiehl}
\author[aff1]{Björn~Lindequist}
\author[aff1]{Nora~Charlotte~Pohlan}
\author[aff6]{Sarah~Schmell}
\author[aff1]{Klaus~Strohmenger}
\author[aff6]{Falk~Zakrzewski}
\author[aff7]{Markus~Plass}
\author[aff8]{Michael~Takla}
\author[aff3]{Tobias~Küster}
\author[aff2]{André~Homeyer}
\author[aff1]{Peter~Hufnagl}

\address[aff1]{Charité -- Universitätsmedizin Berlin, corporate member of Freie Universität Berlin and Humboldt Universität zu Berlin, Institute of Pathology, Charitéplatz 1, 10117 Berlin, Germany}
\address[aff2]{Fraunhofer Institute for Digital Medicine MEVIS, Max-von-Laue-Straße 2, 28359 Bremen, Germany}
\address[aff3]{Technische Universität Berlin, DAI-Labor, Ernst-Reuter-Platz 7, 10587 Berlin, Germany}
\address[aff4]{QuIP GmbH, Reinhardtstraße 1, 10117 Berlin, Germany}
\address[aff5]{Institute of Pathology, University Hospital RWTH Aachen, Pauwelsstraße 30, 52074 Aachen, Germany}
\address[aff6]{Institute of Pathology, Carl Gustav Carus University Hospital Dresden (UKD), TU Dresden (TUD), Fetscherstrasse 74, 01307 Dresden, Germany}
\address[aff7]{Medical University of Graz, Diagnostic and Research Center for Molecular BioMedicine, Diagnostic \& Research Institute of Pathology, Neue~Stiftingtalstrasse~6, 8010 Graz, Austria}
\address[aff8]{vitasystems GmbH, Gottlieb-Daimler-Straße 8, 68165 Mannheim, Germany}

\cortext[corresp]{Corresponding author}

\begin{abstract}
  Over the past decade, artificial intelligence (AI) methods in
  pathology have advanced substantially. However, integration into
  routine clinical practice has been slow due to numerous challenges,
  including technical and regulatory hurdles in translating research
  results into clinical diagnostic products and the lack of
  standardized interfaces.

  The open and vendor-neutral EMPAIA initiative addresses these
  challenges. Here, we provide an overview of EMPAIA's achievements
  and lessons learned. EMPAIA integrates various stakeholders of the
  pathology AI ecosystem, i.e., pathologists, computer scientists, and
  industry. In close collaboration, we developed technical
  interoperability standards, recommendations for AI testing and
  product development, and explainability methods. We implemented the
  modular and open-source EMPAIA platform and successfully integrated
  14 AI-based image analysis apps from 8 different vendors,
  demonstrating how different apps can use a single standardized
  interface. We prioritized requirements and evaluated the use of AI
  in real clinical settings with 14 different pathology laboratories
  in Europe and Asia. In addition to technical developments, we
  created a forum for all stakeholders to share information and
  experiences on digital pathology and AI. Commercial, clinical, and
  academic stakeholders can now adopt EMPAIA's common open-source
  interfaces, providing a unique opportunity for large-scale
  standardization and streamlining of processes.

  Further efforts are needed to effectively and broadly establish AI
  assistance in routine laboratory use. To this end, a sustainable
  infrastructure, the non-profit association EMPAIA International, has
  been established to continue standardization and support broad
  implementation and advocacy for an AI-assisted digital pathology
  future.
\end{abstract}


\end{frontmatter}

\section{Introduction}

Advances in precision medicine, enabled by improved diagnostic accuracy
and novel therapeutics, have significantly improved clinical outcomes
for patients. Pathology, a central diagnostic part of precision
medicine, thus faces a substantial increase in
workload~\cite{BonZafMau2021, WarSteAnd2016, WolPhiLah2023}. At the same time, there is a global
shortage of pathologists~\cite{RobGupCra2015, MetColLeu2019, MarFuzHus2021}. AI-assisted image and
data analysis could increase pathologists'
productivity~\cite{HanReuSam2019, BaiBucLee2018, RetAneMor2020, LujQuiHar2021}. Researchers have published many
promising algorithmic solutions~\cite{BaxEdwMon2022, RodRodSil2022}. However, the
path to wide clinical adoption is difficult. A core problem is a lack of
standardization and interoperability for the seamless integration of
image analysis methods into diverse image management and laboratory
information systems. Commercialization and clinical implementation of
pathology AI must overcome additional hurdles~\cite{Huf2021, HomLotSch2021},
namely the transformation of an idea into an AI prototype (which
requires data acquisition), a validation process towards market
readiness, and certification as a medical product. Finally,
reimbursement and billing issues must be solved to generate revenue.

The EcosysteM for Pathology diagnostics with AI Assistance (EMPAIA)
consortium was established in response to the 2019 innovation
competition ``Artificial Intelligence as a Driver for Economically
Relevant Ecosystems'' by the German Federal Ministry for Economic
Affairs and Climate Action. EMPAIA's mission was to promote the
ecosystem for AI in pathology by involving all relevant stakeholders and
addressing issues of digitalization, standardization, legal and
regulatory requirements, and billing. In addition, the sister project
``EMPAIA Austria'' was funded by the Austrian Research Promotion Agency
to conduct related research on user interfaces and explainability.

The specific objectives of EMPAIA were:

\begin{enumerate}
\item Specify open interfaces for interoperable AI apps in pathology,
  taking into account requirements of different vendors.
\item Collaborate with diverse pathology labs (``reference centers'')
  to evaluate and obtain practical feedback on the EMPAIA
  developments.
\item Support AI vendors by open-source reference implementations of
  the interfaces and by recommendations on product development and
  regulatory affairs.
\item Develop explainable AI (XAI) approaches specifically for
  pathologists.
\item Transfer knowledge and facilitate exchange between all
  stakeholders.
\end{enumerate}

\begin{figure*}[t]
  \centering
  \includegraphics[width=0.9\textwidth]{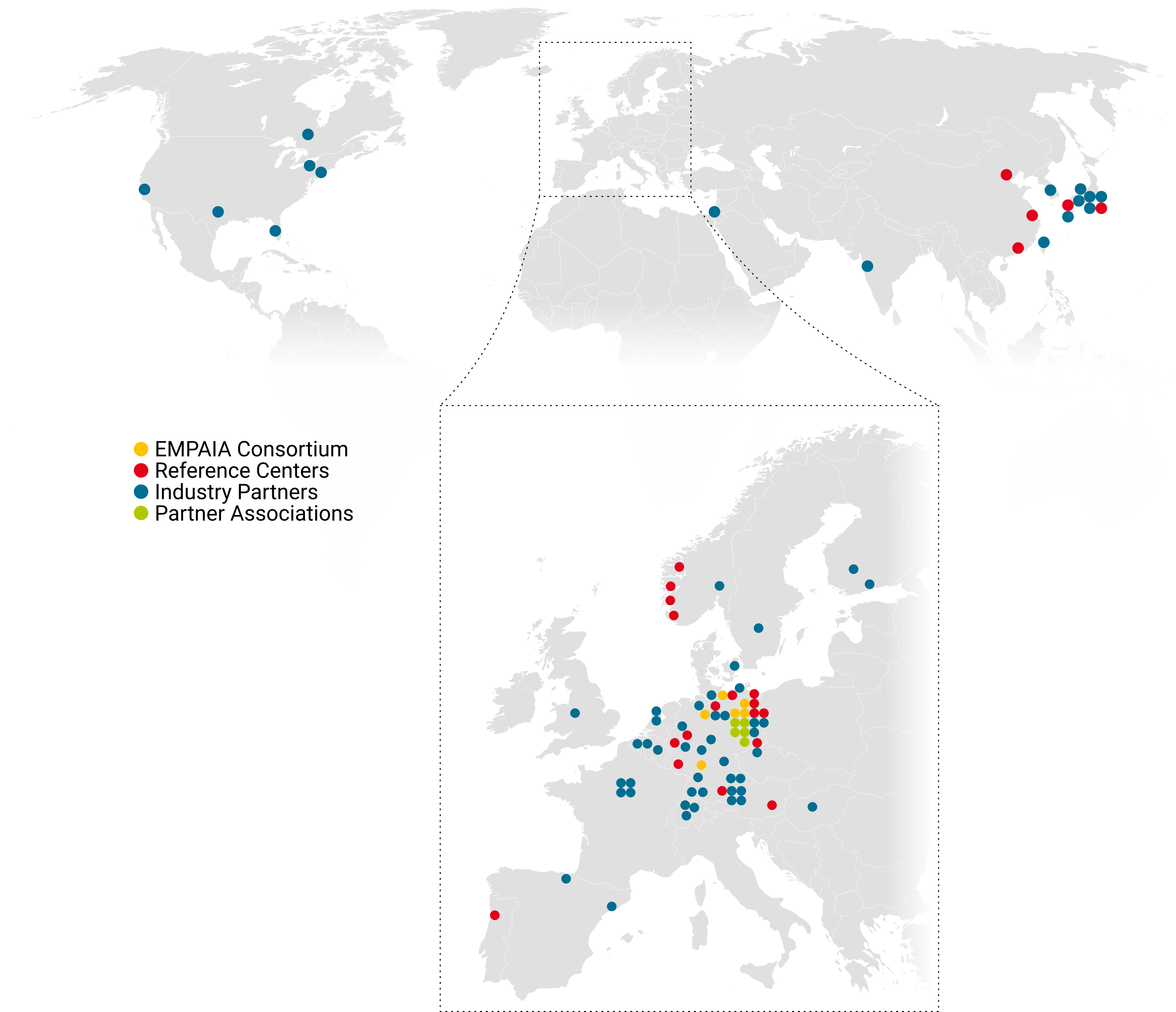}

  \caption{Locations of national and international EMPAIA partners.
  (World map adapted from Wikimedia Commons~\cite{WorldMap}).}
  \label{fig:map}
\end{figure*}

The core EMPAIA team comprised 25 members, including pathologists,
informaticians, mathematicians, technicians, designers and supporting
staff. Many international stakeholders have been involved from the
beginning, including vendors of AI solutions/pathology software
systems/scanners, diagnosticians from large and small clinical
institutions, pharmaceutical companies, and AI researchers. Having
started with 26 commercial partners and 5 partner associations, the
network has grown substantially to 59 commercial partners,
5~associations, and 16 reference centers (Fig.\,\ref{fig:map}). An updated list of
consortium members, industry partners, and reference centers can be
found on the project webpage~\cite{EMPAIAWeb}.

Here, we discuss the achievements of EMPAIA, including the corresponding
lessons learned and their future continuation. We describe technical
developments, such as the specification of unified interfaces for AI
apps together with multiple industry partners and how the EMPAIA
platform can facilitate their adoption. In this context, we also
describe how we collaborated with reference centers to test the
developments in real workflow scenarios. We summarize our research
findings for XAI in pathology to make analysis results more
understandable and gain user acceptance. We describe how EMPAIA
supported AI providers with regulatory issues, for instance, by
publishing guidelines~\cite{HomGeiSch2022} and initiating a service for
the validation of AI solutions. We describe how we reached out to
diverse audiences of the digital pathology ecosystem to increase
visibility and engage relevant stakeholders. Lastly, we provide an
overview of synergies with related initiatives and an outlook on how the
newly-founded non-profit EMPAIA International association will drive
forward these activities in the future.

\section{Technical Developments and Evaluation}\label{sec:technicalDevelopments}

\subsection{Standardization}

For efficient use in the diagnostic workflow, AI solutions must interact
with various systems in the laboratory's IT infrastructure. The basis of
this infrastructure is an AP-LIS, which manages case and sample
information, including diagnostic results. An IMS handles slide scans
for whole slide images (WSIs), which can be implemented as a picture
archiving and communication system (PACS) or vendor-neutral archive
(VNA)~\cite{HomLotSch2021}. In addition, there are pathology workstations
with graphical user interfaces for WSI viewing and interacting with AI
solutions, compute infrastructure for high-performance AI processing, AI
application registries, and billing systems.

The use of AI solutions is severely hampered by a lack of
interoperability~\cite{RomStrJan2022}. Some software systems already
offer their own application programming interfaces (APIs) for executing
AI solutions and exchanging input and output data. In this process, WSIs
are usually represented according to proprietary file format
specifications from various scanner vendors. AI developers must adapt
their solutions to all these different APIs and WSI formats, leading to
repetitive integration work and delays in the availability of AI
applications, which drives up their costs. Many systems currently do not
even offer any interfaces for AI solutions~\cite{EscCarBoc2022}.

\subsubsection{The EMPAIA Platform}

To improve this situation, we specified the EMPAIA App Interface, an
open and vendor-neutral standard for integrating AI solutions into
pathology software systems~\cite{RomStrJan2022}. The API was developed in
close cooperation with associated industry partners, specifically
considering their requirements. To facilitate the adoption of the EMPAIA
App Interface, we published comprehensive documentation describing the
technical implementation from the perspective of developers of AI apps
and pathology software systems~\cite{DevWeb}. In addition, we
released the EMPAIA App Test Suite, an open-source toolbox for
automatically checking the compliance of a given AI
app~\cite{EATSgitlab}. With the documentation and test suite,
third-party developers were able to integrate and test their
applications without further advice.

\begin{figure*}[t]
  \centering
  \includegraphics[width=0.7\textwidth]{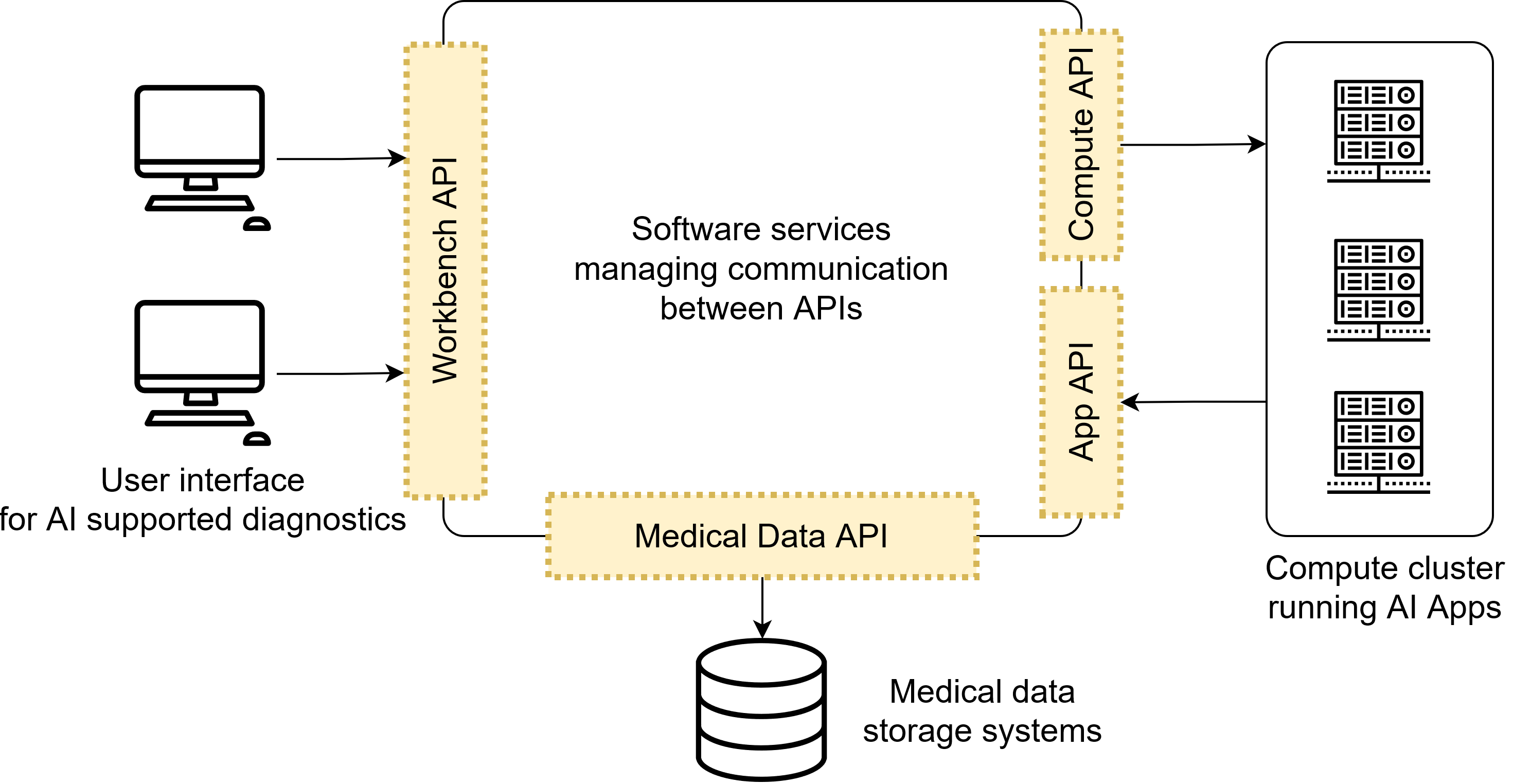}

  \caption{EMPAIA Platform architecture. Simplified diagram of the
    EMPAIA Platform architecture, showing the main APIs serving as
    abstractions over clinical system software components.}
  \label{fig:architecture}
\end{figure*}

The EMPAIA App Interface is embedded in the EMPAIA Platform (Fig.\,\ref{fig:architecture}).
Its highly modular architecture demonstrates how the different systems
of the laboratory IT infrastructure can be integrated effectively to
enable the use of AI~\cite{JanLinStr2023}. The platform provides
ready-to-use implementations of critical components, such as a user
workbench with a virtual microscope viewer and data management services.
The platform is purely prototypical, neither certified as a medical
product nor intended to compete with commercial pathology software
systems. However, it is open source and subject to the industry-friendly
MIT license~\cite{MITWeb}, so both software system providers and
laboratory IT administrators can use the platform as a blueprint and
testbed for custom implementations.

The EMPAIA Platform comprises several APIs that represent abstractions
of different software systems (Fig.\,\ref{fig:architecture}). These abstractions allow for
flexibility in the implementation of the technical components and enable
combining different programming languages and database technologies
while maintaining compatibility between the systems. The heart of each
AI app is a model that takes image data and other parameters as input,
processes the image and produces output data. EMPAIA specifies how such
AI models are encapsulated using container technologies for uniform
execution. AI~models communicate via the EMPAIA App Interface to
transmit input and output data. Equally important are user interfaces
that allow pathologists to view WSI, interact with the AI solution,
e.g., by drawing a region of interest (ROI), and render processing
results. The EMPAIA App UI (user interface) concept allows AI vendors to
bundle a custom web-based user interface with their apps, which can
provide an optimized user experience for the specific diagnostic task.
It also potentially reduces the effort required for regulatory
compliance when integrating into new systems because AI models are
usually validated in the context of the full diagnostic workflow. This
diagnostic workflow includes the graphical user interface that displays
AI~results in a specific way and, depending on the AI~app, might enable
user interactions like drawing an ROI or selecting hotspots for further
analysis. EMPAIA allows custom app user interfaces to be seamlessly
embedded into browser-based pathology workstations and to connect with
the underlying clinical system via the EMPAIA Workbench
API~\cite{DevWeb}.

\subsubsection{Lessons Learned}

Since the utility of the EMPAIA Platform is highly dependent on the
acceptance and contributions of external AI companies, it was essential
to pursue an agile development approach. First, a detailed requirements
analysis was conducted together with AI companies to identify a small
subset of features covering the minimum requirements of many products on
the market. This allowed us to quickly release a minimum viable version
of the platform. While this version was limited to the analysis of ROIs
drawn by users in a WSI viewer, it enabled AI developers to create first
usable prototypes of their apps and to provide early feedback for future
development. Building on the minimum viable version, the platform was
iteratively expanded and improved, for instance, by further annotation
types, configuration options, or preprocessing functionality. The
development process also revealed previously unknown requirements. In
particular, the need to enable custom app user interfaces was only
identified through feedback from regulatory experts. It made our
original concept of a unified user interface for all apps
obsolete~\cite{JanLinStr2023}.

The inconsistent implementation of apps from different vendors proved to
be particularly challenging. On the one hand, the apps were implemented
based on different computing platforms and paradigms, e.g., as
Windows-based client applications or Linux-based cloud applications.
Only container technologies allowed all apps to be integrated uniformly
into the EMPAIA Platform. On the other hand, the apps expected different
data inputs: While some apps could open WSIs of different formats by
themselves, others depended on corresponding functionality in the host
platform. Therefore, the platform had to be highly modular to supplement
missing functionality as needed. This requirement is fulfilled with a
layered design of APIs that serve as abstractions of the underlying
implementations. Furthermore, the reference implementation follows a
microservice architecture that allows parts of the implementation to be
replaced, updated or improved as necessary, without affecting unrelated
parts of the platform.

\subsubsection{Perspective}

The EMPAIA Platform is continuously being extended to new requirements.
To coordinate future developments, we set up the EMPAIA Community Group,
an open communication forum for all stakeholders~\cite{CommunityWeb}.
Changes and extensions to the platform are discussed in a transparent
and formal manner through EMPAIA Mod Proposals~\cite{ModpropWeb}. One
of the most important current proposals is adopting DICOMweb (see
below). Another highly requested extension is the possibility of storing
and visualizing pixel-wise data overlays on WSIs~\cite{Modprop102Web}.
Such extensions would improve communication with DICOM-based PACS
systems and enable AI apps to provide explainability functionality, thus
simplifying integration and increasing the adoption of AI in clinical
practice.

Clinical systems (AP-LIS, IMS) currently used in practice still have to
work with various proprietary WSI file formats because the DICOM WSI
standard has yet to be widely adopted. The EMPAIA team designed a common
API that serves as an abstraction of such file formats (including DICOM
WSI files). This approach allowed for a quick start in development to
deliver a practical solution but resulted in an API that is not
specifically optimized for DICOM. Since DICOM is widely considered to be
the future standard for WSI storage and
transmission~\cite{EscCarBoc2022}, the EMPAIA team is investigating a
shift towards the DICOMweb standard on the API level as part of an
upcoming platform release while still supporting the usage of
proprietary formats in the backend~\cite{Modprop101Web}.

The IHE PaLM Technical Committee provides guidelines for Pathology and
Laboratory Medicine (PaLM)~\cite{DasJonMer2021}. For instance, the
committee published the Digital Pathology Image Acquisition (DPIA)
profile~\cite{IHEPaLMWeb}, describing the digitization and storage
process for WSIs. The EMPAIA specifications complement the IHE proposals
by specifying how AI apps can access image data after storage in a
clinical system.

Adopting the open EMPAIA App Interface and building on the EMPAIA
Platform is in the hands of the industry. As of Dec. 2023, 14 AI apps
from eight companies have been integrated into the EMPAIA
platform~\cite{PortalWeb}, and more integrations are being actively
developed. Four international vendors of pathology software systems are
in the process of enabling the APIs in their products. A growing catalog
of compatible AI apps has the potential to encourage more software
system providers to undertake integration efforts to enable their
customers to use such apps, which, in turn, will trigger network effects
that will encourage more AI companies to join the initiative. The EMPAIA
specifications impose strict requirements and a separation of concerns
between clinical systems and AI apps. EMPAIA, therefore, offers the
advantage of a unified approach for interoperability, reducing the
multiplied efforts of integrating many heterogeneous AI apps of
different companies with various clinical systems.

\subsection{Reference Centers}

For an ecosystem project, it was crucial to partner with a variety of
laboratories that used apps via EMPAIA. These partnerships enabled a
better understanding of the requirements for adopting and using AI in
clinical routine, testing the platform in practical settings,
identifying and understanding issues, and discussing the potential for
improvement.

For this reason, EMPAIA collaborated with multiple national and
international reference centers, including seven university hospitals,
two hospital chains and six larger private practice groups. The
reference centers implemented and tested the EMPAIA platform on-site and
provided feedback from a clinical/technical perspective. They also
participated in a user study evaluating the image analysis apps
available via the platform. In addition, the reference centers performed
their own research in the context of EMPAIA~\cite{SerLueRod2023, KerBulKli2022, HolBouJoo2023, EchLalQui2022, VafBulSti2023}. The
centers were at very different stages of the digitization process, which
created additional challenges but was crucial for accurately
representing real-world conditions.

\subsubsection{Platform Deployment and Usage}

To evaluate the EMPAIA platform, reference centers were provided either
access to a deployment in the EMPAIA cloud or an on-premises offline
deployment. The deployment included a prototypical data management
application that is part of the EMPAIA platform. This way, the
evaluation could begin even before a full LIS and IMS integration was
completed. The data management application performed client-side
anonymization before data transfer, mitigating data protection concerns
and avoiding unnecessary transfer of patient data~\cite{BisFraDog2023}.
For the evaluation, the reference centers could run selected analyses on
demand, i.e., by drawing ROIs to trigger computations. The example apps
available comprised a mix of research use-only, non-approved, and
CE-IVD-approved apps. For legal and licensing reasons, app usage had to
be restricted to evaluation and clinical use was prohibited.

To familiarize pathologists and qualified staff with AI at the reference
centers, we asked them to evaluate multiple apps integrated into the
EMPAIA platform. In this context, we collected structured feedback from
these users on the perceived correctness of histopathological tissue
analysis with AI-assisted assessment compared to human-only assessment
and feedback about usability and bugs. Pathologists, medical students,
and researchers, including users with extensive routine experience in
pathology and digital natives, were included to cover a diverse range of
users. Depending on the preferences of the reference centers, human-only
assessment was performed using a routine microscope or pathology viewer
software. This collection of user feedback focused on obtaining
actionable feedback for iterative improvement and thus provides
anecdotal results rather than unbiased quantitative performance data.

\subsubsection{Lessons Learned}

Despite obvious advantages, such as reducing efforts in procuring
hardware or deploying in-house infrastructure, using cloud services
posed a significant problem for some institutions due to IT security
vulnerability and privacy concerns. For cloud-based deployments, a
privacy agreement had to be reached between reference centers and
EMPAIA, ensuring the secure transfer of anonymized medical data. As some
apps depend on external cloud services, such apps were unavailable in
the on-premises deployments. This indicates a lose-lose situation known
from the ongoing controversy over local vs. cloud solutions: vendors of
cloud-only apps lose potential clients, while labs using only offline
apps lose potentially valuable tools.

User feedback provided clues as to why satisfaction with app assistance
still has room for improvement. Besides issues with image quality, which
need to be avoided by optimized lab workflows, and outages of the EMPAIA
infrastructure, feedback concerned issues concerning app results, app
usability and app usage as part of the current platform. After
categorization, we clarified details with the reference centers, curated
the feedback, and forwarded it, as appropriate, within EMPAIA or to the
app developers for improvements.

One app yielded results that were obvious to be inadequate in several
labs, and another yielded inadequate results in a single lab, possibly
due to the incompatibility of WSIs generated with a specific scanner.
One app was criticized for not yet computing diagnostically relevant
quantities from the image analysis results, and one app was criticized
for poor usability. On the other hand, one app was praised for its
capability to detect and quantify weak signals barely visible to the
human eye. These issues and feature requests were followed up with the
vendors.

The process of drawing an ROI and waiting for the computation of
analyses was often perceived as excessively slow, which occasionally
compelled users to create ROIs that were too small for any meaningful
analysis. This underlines the need for unobtrusively fast
implementations and, e.g., running computationally expensive WSI-wide
analyses before a pathologist starts interacting with the case. The
latest EMPAIA API version supports such preprocessing for compatible
apps. Manual correction of algorithmic results was viewed as cumbersome.
App results that did not align with criteria present in pathological
analyses comprehensible and assessable by pathologists (e.g., glandular
units) were not trusted, underlining the necessity of suitable
explainable AI approaches. This indicates the need for more
communication and a collaborative app development process between
developers and users. This should lead to a clearer alignment of user
expectations, app capabilities, and presentation of algorithmic results.
Moreover, users considered integration with LIS indispensable and this
issue will need to be evaluated once available. Operating apps as
isolated tools was not considered useful for routine diagnostics.

These findings provide valuable user feedback both for the improvement
of individual apps and for the next steps of integration in an iterative
manner. The evaluation so far has focused on single apps. Once
combinations of apps are available to assess complete clinical cases
with different stainings, usability studies will need to investigate to
which extent user interaction should be homogenized across apps by
different vendors. App performance should be evaluated in more detail,
considering real patient cases with slides of different staining, i.e.,
requiring a targeted study setup for app validation. For this purpose,
providing multiple apps via the same platform permits head-to-head
comparisons and allows users to choose the optimal apps for their
requirements.

\section{Explainability}

Apps using AI algorithms are typically intransparent with respect to how
results are obtained from the input images. There is an agreement on the
need to control and understand AI solutions, especially when being used
in potentially life-threatening domains such as medical
diagnostics~\cite{TrustworthyAIReport, AIRegProposal}. However, there are multiple facets
to explainability, e.g., which aspects should be explained to whom.
Currently, no consensus exists on terminology and definitions of
explainability of AI.

In EMPAIA, an AI application typically performs deep learning-based
image analysis on WSIs. It, therefore, consists of one or more deep
neural networks (models) and additional algorithms for pre-,
intermediate and post-processing tasks. While such deep learning models
can achieve very high task performance (e.g., classification accuracy,
intersection over union for segmentation tasks), model size and
complexity are barriers to understanding their inner workings, making
them a ``back box.'' Since their high-dimensional decision space and
stochastic origin provide no inherent guarantee for correctness, one
needs other means of estimating their function and reliability. Since
the qualities of an explanation are highly dependent on the recipient of
the explanation, the situation in which it is given and the goal it is
expected to support, one needs to define this context in order to
effectively discuss, select and measure an explainability
method~\cite{ShaSivAlb2022}.

We systematically investigated aspects of explainability within and
around the context of using AI-based diagnostic support software in a
clinical setting~\cite{EvaRetGei2022}. This includes detailed knowledge
of the stakeholders to provide customized
explainability~\cite{PlaKarKie2023}. We address different stakeholders,
with a focus on operators of the AI software (pathologists), developers
of AI software, decision-makers, and validators who need to make
high-level decisions about the use of AI software. In the following
paragraphs, we will lay out their use cases, goals and our efforts
towards providing explanations for each of these use cases.

The first and most prominent user group are the pathologists or
technical assistants who are directly operating the AI application
within tight time restrictions and with the goal of providing the
correct medical diagnosis. In this group, explainability is a potential
source of trust and safety. As pathologists are the ones who have to
decide and are responsible for their decisions, they have an inherent
need for information that helps them with their judgment.

We surveyed pathologists to learn about their general preferences and
understandings regarding explainability methods and
explanations~\cite{EvaRetGei2022}. Besides favoring explanations that are
close in style and content to what another pathologist would provide
them with, we also noticed that some misinterpreted the explanations for
segmentations. Besides the machine learning-related part of XAI, we also
looked at human-centric evaluations. We found that user-friendly
interfaces, enabling pathologists to explore causal relationships, play
an essential role in keeping the human in the loop. However, such
interfaces require more research and innovation~\cite{PlaKarKie2023}.
\looseness-1

\begin{figure*}
  \centering
  \includegraphics[width=0.7\textwidth]{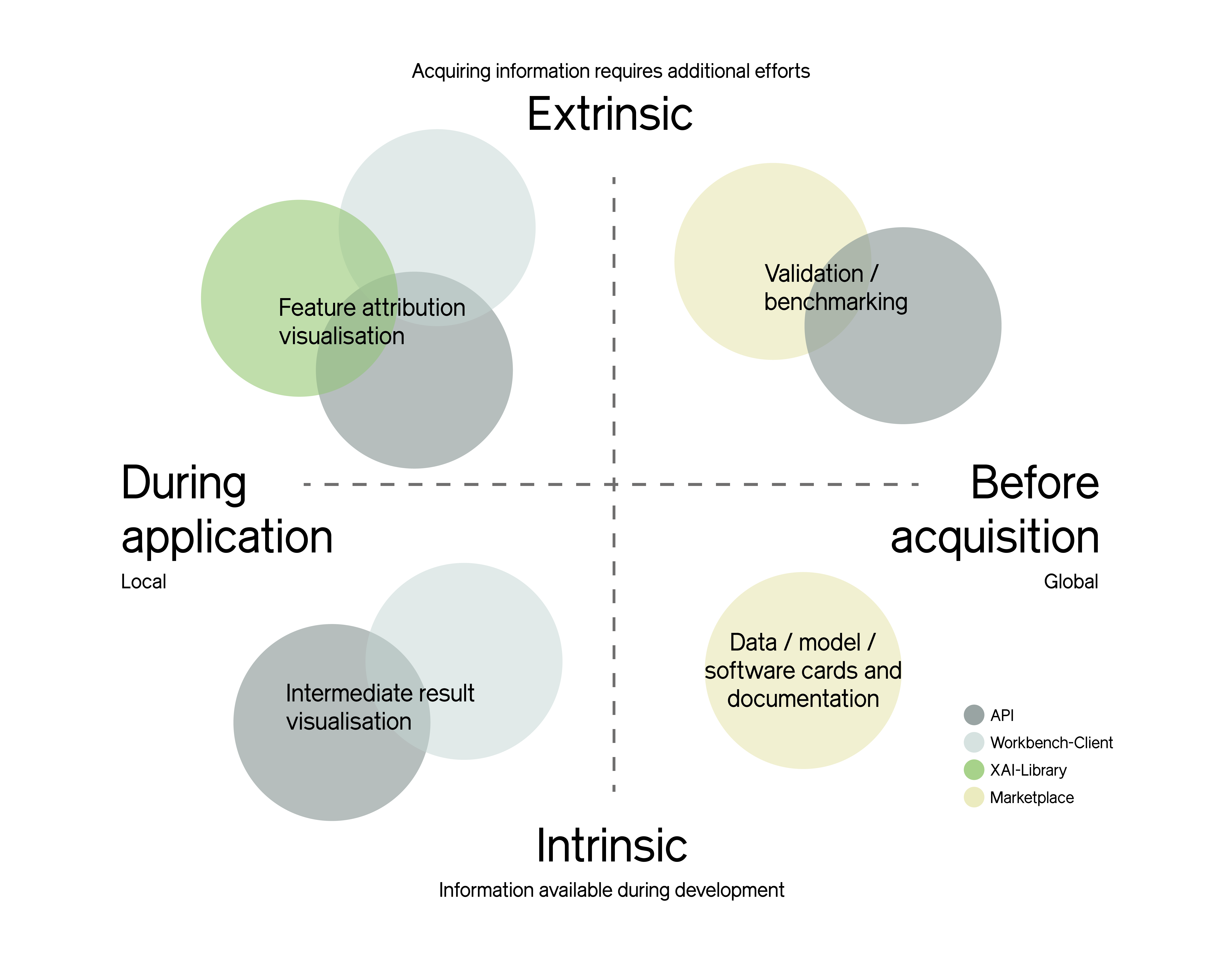}

  \caption{XAI Approaches. Overview of different XAI approaches, the
    level they address and the project deliverables that have been
    affected or designed to support them. Local approaches provide
    explanations during usage of AI applications and directly address
    pathologists. Global approaches are supplements of AI applications
    that help decide whether they should be acquired in general.}
  \label{fig:XAI}
\end{figure*}

Pathologists require explanations on a specific slide, sample or case.
These instance-based explanations are called local explanations. To
provide XAI on the pathologists' desk, it has to be available and easy
to use, e.g., by integrating it directly into the applications and user
interfaces they use in their daily work, alongside the actual results of
the ML algorithms. This can be achieved in many ways (Fig.\,\ref{fig:XAI}), with
varying degrees of explanatory value and understandability or the
potential for misinterpretation if unfamiliar with the underlying
method.

The second stakeholder and use case that we address are application
developers. They aim to provide trustworthy, e.g., reliable,
well-performing applications that follow legal and ethical regulations.
Explainability methods can play a crucial role in reaching these goals.
On the regulatory side, over the last few years, the EU published a set
of directives and drafts for guidelines that already motivate and will
create a strong demand for explainable AI. On the technical side, deep
learning works stochastically and by example; if there were any
unexpected and unwanted correlations (bias) in the training data,
developers need ways to find out about such effects in order to address
them.

Besides surveys, we investigated existing explainability methods
applicable on image-based deep learning models. We developed a
state-of-the-art model-agnostic explainability framework, including
efficient optimization algorithms that speed up many existing
model-agnostic methods by drastically reducing sampling
efforts~\cite{ShaSivAlb2022, ShaAlb2022}. We released these results together with
implementations for several sampling-based XAI methods as an open-source
library~\cite{MAXLibCode}. Model-agnostic methods can be used to
generate feature attribution heatmaps in cases without access to the
model (external evaluation), or the model itself is not differentiable.
Another reason can be the case of introducing new model architectures or
transfer functions for which no reversible approaches exist yet.
Model-agnostic methods come at a higher cost of execution. Still, they
are cheaper to adapt to many different models, so they can help
mass-validate vast ranges of different ML models when evaluating them
before use in practical applications.

When several AI applications are run next to each other, consistent
heatmap color gradients across apps should be used to avoid confusing
users. To make the results of explainability methods, such as
sensitivity maps, customizable, we created concepts and a software
prototype for a standard of transmitting explainability information in a
resolution and type similar to the original WSI. A crucial sub-concept
in this API proposal~\cite{Modprop102Web} was the distinction between
content and its visual representation. As our previous observations
highlighted, there is a need for customizable visualization depending on
the shown information, stain color of the underlying WSI, etc. The
second motivation are standardized user interfaces to enable the
customizability mentioned above and consistency regarding color
gradients. To realize this, our standard transmits only physically
meaningful values together with semantic information required to select
the appropriate color scheme to map to. Once implemented, these concepts
must be verified in a usability study.

Finally, the third stakeholders we addressed are those who decide
whether an AI application will be acquired or is safe to use in medical
practice, e.g., hospital administrators, purchasing managers, and
regulators. They typically require global explanations and information
about how well an AI application performs on all of the patients rather
than a single patient. This can also include information about the data
that was being used for training the underlying models or the data that
has been used to evaluate the models. Typically, this information is
collected in model cards and can be shown in a portal like the prototype
we developed in EMPAIA. A reliable and well-performed validation is the
basis for supporting these stakeholders with such global explanations.

\section{Regulatory Perspective}

Apps in pathology used for diagnostic or therapeutic purposes are
subject to the In Vitro Diagnostic Medical Device Regulation
(IVDR)~\cite{EU2017Reg746} in the EU and the Code of Federal
Regulations~\cite{ECFRWeb} in the USA. Meeting these regulatory
requirements is a major challenge when building AI products in
pathology. Additional challenges include software quality, integration
in laboratory IT infrastructure, business models, and reimbursement. To
provide software vendors with advice on how to meet these challenges, we
conducted extensive research and published an open guidance
paper~\cite{HomLotSch2021}. In addition, EMPAIA is monitoring the
pathology AI apps available on the market and their regulatory approval
status in different jurisdictions and provides this directory
online~\cite{ClearancesWeb}.

To demonstrate the practical utility and obtain regulatory approval, AI
apps must be validated appropriately~\cite{EU2017Reg746}. A major
challenge in validating AI apps is compiling suitable test datasets.
These must be sufficiently diverse to cover the extensive biological and
technical variability of WSIs. They must also be sufficiently large to
obtain statistically meaningful performance estimates. Moreover, test
datasets must cover relevant subsets, be unbiased, and be sufficiently
independent of datasets used for development.

In the past, there was little guidance available on how to compile test
data sets, which was a major barrier to adopting AI solutions in
laboratory practice. To change this, we organized the committee
``Validation of AI Solutions,'' consisting of more than 40
representatives from different stakeholder groups of the EMPAIA
ecosystem, including pathologists, AI vendors, pathology software system
vendors, and scanner vendors. The committee met in regular video
conferences to discuss and define recommendations for creating and using
test datasets based on literature research and experience. The results
were published as an open guidance paper for AI developers and
pathologists~\cite{HomGeiSch2022}.

To ensure that performance estimates are unbiased, AI apps must be
tested on datasets that are independent of their training data and
originate from different clinics and patients. Testing should ideally be
conducted by an independent body with no conflicts of interest. For this
reason, the EMPAIA App Validation Service was initiated in late 2022. AI
vendors can submit their solutions to this service to have them tested
against independent datasets curated by EMPAIA according to the proposed
guidelines~\cite{HomGeiSch2022}.

In a first prototypical phase, the service provides validation of AI
apps for PD-L1 immunohistochemistry (IHC) on WSIs in non-small cell lung
cancer (NSCLC), detecting positive and negative tumor cells and
calculating the Tumor Proportion Score (TPS). The dataset used for this
purpose consists of over 200 slides from 22 cases that were stained with
eight different antibodies. Reference TPS were provided by panels of
three institutes. All slides were digitized using five different
scanners, resulting in over 1000 WSIs. AI vendors do not receive the
validation dataset itself, as sharing the dataset would no longer ensure
an unbiased performance assessment for future AI solution submissions.
Instead, AI vendors make their solutions compatible with the open EMPAIA
Platform infrastructure~\cite{JanLinStr2023}, so that testing can be
carried out autonomously by the EMPAIA App Validation Service. As a
result, vendors are provided with a report of the specific strengths and
shortcomings of their apps with regard to different tissue regions,
stains, and scanner types.

\section{Communication and Knowledge Transfer}

\subsection{Public Relations}

For an ecosystem project, it is crucial to be visible to find additional
partners, in our case app developers, LIS developers, and laboratories,
and to attract the attention of future end-users, i.e., pathologists and
decision-makers. These primary target groups are a diverse audience with
backgrounds in computer science, management, and medicine. Five main
channels were used to disseminate information: website, newsletter,
social media, peer-reviewed publications, and conferences. Project news
included events, publications, API updates, software, documentation
releases, and the availability of EMPAIA Academy material.

Besides communication to the general public, we also disseminated
project results to the scientific community in articles providing a
project overview~\cite{Huf2021}, describing aspects implemented in
the EMPAIA platform~\cite{RomStrJan2022, BisFraDog2023, JanStrRom2022}, giving recommendations to
different stakeholders~\cite{HomLotSch2021, HomGeiSch2022, JanLinStr2023, SchKieCar2023}, and presenting
other EMPAIA-related research~\cite{ShaSivAlb2022, EvaRetGei2022, PlaKarKie2023, ShaAlb2022, MulKarPla2022, PlaKarEva2022, SchSchGei2022, Kie2022}.
Furthermore, congresses and events worldwide presented opportunities to
showcase EMPAIA. These events allowed talking to media representatives,
pathologists, and AI experts and to gain visibility and exchange
information about the project and current developments in the field of
digital pathology to establish and intensify networks.

\subsection{EMPAIA Academy}

Activities involving stakeholders to deepen their shared knowledge of
digital pathology were deemed crucial since the initiative intends to
link people and software systems. We have grouped these programs
together under the name ``EMPAIA Academy''. We provided medical experts
with an understanding of the basics of data science and AI development.
Conversely, we showed AI developers the challenges of quality
diagnostics and the legal considerations involved in creating advanced
software and interfaces for a medical field. The EMPAIA Academy offered
two types of workshops: lectures by selected invited domain experts and
hands-on sessions at conferences, which provided insight into the
technical activities behind data preparation and training of machine
learning models.

Two rounds of half-day courses, IT/AI for pathologists and pathology for
IT/AI specialists, were delivered as online webinars with free
registration and recordings made available online to facilitate
attendance. Basic courses covered laboratory workflow, pathologists'
diagnostic work, introduction to WSI scanning, examples of AI
applications in pathology, and best practices for algorithm development
(for IT/AI specialists), as well as an introduction to machine learning,
application of such techniques in pathology, and examples of AI use in
different subspecialties (for pathologists). Advanced courses covered
the productization of AI algorithms, an overview of existing products,
and the regulatory and legal challenges of using AI in pathology. The
hands-on sessions were held as pre-conference workshops at the European
Congress on Digital Pathology in 2022 and 2023 and material was made
available online afterwards~\cite{ColabMitosisWeb, ColabTilingWeb, ColabIntroWeb}. They consisted of
several parallel tutorials, each covering different facets of
computational pathology. Designed as activities requiring little prior
knowledge of programming or pathology, the tutorials included
interactive analysis of WSIs, automated image processing techniques,
approaches for patch-based image analysis, and AI-based image
classification. During the workshops, collaboration and discussions in
small groups led to a lively atmosphere and networking. These workshops
also stood out from the industry pre-conference workshops in that they
provided knowledge to participants with no commercial interest.

\section{Discussion}

Expectations for AI support in diagnostic pathology are high but are
currently not being met. This is due to technical and regulatory
challenges in translating research results into clinical diagnostic
products and the lack of standardized interfaces between systems, among
other reasons. We aimed to overcome these hurdles by developing standard
interfaces for integrating AI apps into the workflow and by using these
apps in routine clinical diagnostics, by supporting the validation and
certification of AI apps, by fostering the exchange of knowledge between
stakeholders (pathologists, algorithm developers, vendors, scanner
vendors, and AI app vendors), and by improving the explainability of the
results.

Starting from a digital pathology landscape with only a few proprietary
interfaces to connect individual pathology workstations or IMS with AI
apps from selected vendors, EMPAIA created the first open and
vendor-independent interface specification. This was subsequently
adopted by multiple vendors and tested in reference centers with
encouraging results, providing a proof of concept of how different AI
solutions can be integrated in a standardized way. The EMPAIA interface
specification is being continuously extended to better integrate AI
solutions into the laboratory workflow, e.g., by further improving DICOM
interoperability and supporting different licensing models for billing
(e.g., cloud-based and on-premises, time-based and usage-based).

By collaborating with reference centers, a number of pathologists were
able to familiarize themselves and gain experience with AI. In addition,
many pathologists benefited from the knowledge transfer at the EMPAIA
Academy. Similar to previous introductions of new methods in pathology
(e.g., IHC, fluorescence in situ hybridization, molecular diagnostics),
it will be important to expand training programs for clinical users in
cooperation with pathologists' committees and organizations.

We have emphasized our position as an open and vendor-independent
ecosystem by making the educational materials freely available and
releasing the software developed for the EMPAIA platform under the
industry-friendly MIT open-source license. As a side-effect, this also
allows the reuse of components and the adoption of interfaces beyond the
intended purpose of making commercial AI apps available for routine
pathology. It is conceivable that our developments could be used to
create a platform for automated comparison of algorithms, to enable
side-by-side comparison of apps in terms of quality of results,
computational performance, and user experience, and to provide access to
research apps, e.g., for clinical studies.

Obtaining FDA or IVDR approval for AI apps in DP is a major hurdle,
especially for small vendors. This is sometimes circumvented by
marketing apps for research use only, making it even more important to
support the steps necessary for clinical use. While the first AI
applications have been certified by the FDA or according to the IVDR in
recent years, it is still not very clear what specific requirements
vendors must meet. Therefore, we aimed to remove uncertainties about
regulatory processes and, in particular, support the compilation of test
datasets by deriving recommendations from the information available so
far. We expect that the standardization of interfaces, as provided by
EMPAIA, will simplify and accelerate certification. In addition to
recommendations, we have created a first validation dataset
(\textgreater 1000 WSIs from \textgreater 200 slides) as the basis for a
service where AI vendors can obtain independent, external validation of
their image analysis apps.

\subsection{Lessons Learned}

The EMPAIA initiative attracted a great deal of interest from the
industry, as all market participants were facing more or less the same
problems. We succeeded in opening up a pre-competitive space for
interface development. This concept has already been successfully
applied many times, e.g., with DICOM in the medical domain or the USB
interface in consumer electronics. In this process, we underestimated
the persistence with which proprietary approaches are pursued to achieve
market advantages. It took considerable effort, repeatedly emphasizing
that ``we are not selling anything,'' and demonstrating cooperation with
various vendors before EMPAIA was no longer perceived as a potential
competitor.

Setting up the reference centers also proved to be a significant
administrative burden. Hardware acquisition, whether for a research
project or not, must be processed according to the applicable European,
national, and local tendering rules. This, combined with hardware
availability bottlenecks due to the SARS-CoV-2 pandemic and the efforts
of integration in local IT infrastructure, led to very long delays in
establishing the reference centers. In addition, complex data protection
agreements were required. Although no identifying patient data
(including case numbers) was exposed to the image analysis apps, cloud
use was not acceptable to all reference centers. Conversely, not all
vendors provided on-premises implementations of their solutions. These
difficulties are characteristic of the introduction of digital pathology
in connection with AI apps by commercial providers and often result in
stalled or abandoned deployments.

Once installations were established in the reference centers, getting
thorough user feedback proved to be an additional challenge.
Diagnosticians are under a lot of time pressure and are only able and
willing to participate in evaluations, e.g., if the effort is
reasonable. Even when communicated clearly, the fact that the platform,
the apps provided, and the workflow integration are still under
development can lead to a negative overall impression of AI and image
analysis support. Combined with delays in implementation and
improvements, this may also reduce enthusiasm for participating in
further evaluations.

\subsection{Related Initiatives}

EMPAIA is one of several large-scale international initiatives that aim
to support the development and use of AI solutions in pathology or other
medical domains (Table~\ref{tab:relatedInitiatives}). The initiatives complement each other and can
build on each other's work in different ways.

Several initiatives are dedicated to standardization and the creation of
interoperability. The long-established NEMA Medical Imaging Technology
Association (MITA)~\cite{MITAWeb} and IHE PaLM~\cite{IHEWIkiWeb}
initiatives are working on standards for the representation and exchange
of digital pathology images and structured metadata. The best-known
example is the DICOM standard coordinated by NEMA MITA. The EMPAIA
interfaces are already aligned with these standards and compliance is
being further extended to improve the interoperability of input and
output data from AI solutions (see Section~\ref{sec:technicalDevelopments}).
Like EMPAIA, the IHE AI Interest Group for Imaging (AIGI)
Task Force~\cite{IHEEuroWeb} is also developing standards for using
and integrating AI applications into end-user systems, but for radiology
rather than pathology. As this task force started later, it can learn
from the experiences of the EMPAIA project and endeavor to achieve
interoperability between AI solutions in radiology and pathology from
the outset. The joint support of the DICOM standard will facilitate this
process.

The ITU Focus Group on Artificial Intelligence for
Health~\cite{AI4HWeb} (FG-AI4H) is working to create a standardized
assessment framework for AI solutions in medicine, including pathology.
The EMPAIA interfaces can prove useful here, as they make testing AI
solutions from different manufacturers much easier within a standardized
evaluation framework. At the same time, the EMPAIA recommendations for
the creation and use of test datasets can help to obtain meaningful
evaluation results~\cite{HomGeiSch2022}. The ISO Technical Committee 212
is working on the ISO/AWI 24051-2 guidance document~\cite{ISO24051}
on the digitalization and processing of digital whole slide images and
their analysis using AI. The open guidance papers published by
EMPAIA~\cite{HomLotSch2021, SchKieCar2023} can provide a basis for this.

Lastly, multiple large consortium initiatives focus on building
centralized or federated data repositories for pathology images (e.g.,
BIGPICTURE~\cite{BIGPICTUREWEB}, EUCAIM~\cite{ECIIWeb}, NCI Imaging
Data Commons~\cite{FedLonPot2021}, PathLake~\cite{PathLAKEWeb}). By
providing access via the standardized EMPAIA interfaces, these
initiatives can enable the direct application of various AI solutions to
their datasets, e.g., to automatically derive tissue parameters or to
evaluate the AI solution on external data.

\begin{table*}[t]
  \centering
  \caption{Large-scale initiatives that aim to support the development
    and use of AI solutions in pathology or other medical domains
    (alphabetical order; EU: European Union, UK: United Kingdom; DE:
    Germany, USA: United States of America; PF=public funding,
    IF=industry funding; NPO=nonprofit organization)}
  \label{tab:relatedInitiatives}

  \renewcommand{\arraystretch}{1.6}\footnotesize
  \begin{tabular}{L{4.6cm}L{1.4cm}L{1.4cm}L{1.2cm}L{1cm}L{4.1cm}}
    \toprule
    Name                                                                                              & Timeframe            & Country/ Region & No. of partners & Type        & Scope                                                                    \\
    \midrule
    BIGPICTURE~\cite{BIGPICTUREWEB}                                                                   & 2021– ongoing       & EU          & \textgreater 20 & IF, PF      & repository                                                               \\
    EMPAIA / EMPAIA International e.V.~\cite{EMPAIAWeb}                                               & 2019– 2023/ ongoing & DE, global  & \textgreater 20 & IF, PF, NPO & standardization, interoperability, education, AI, regulatory, validation \\
    European Cancer Imaging Initiative (EUCAIM)~\cite{ECIIWeb}                                        & 2023– ongoing       & EU          & \textgreater 20 & PF          & hub, repository, federated analysis (focused on radiology and genomics)  \\
    ITU Focus Group on Artificial Intelligence for Health (FG-AI4H)~\cite{AI4HWeb}                    & 2018– ongoing       & global      & n/a             & PF          & standardization, regulatory                                              \\
    iCAIRD~\cite{iCAIRDWeb}                                                                           & 2018–2023           & UK          & 5–20            & PF, IF      & hub, AI development for radiology and pathology                          \\
    IHE AI Interest Group for Imaging (AIGI) Task Force~\cite{IHEEuroWeb}                             & ongoing             & EU          & n/a             & IF          & interoperability, standardization                                        \\
    IHE PaLM~\cite{IHEWIkiWeb}                                                                        & 2016– ongoing       & EU          & n/a             & IF, NPO     & interoperability                                                         \\
    Imaging Data Commons (IDC)~\cite{IDCWeb}                                                          & ongoing             & USA         & 5–20            & PF          & repository                                                               \\
    ISO (ISO/AWI 24051-2)~\cite{ISO24051}                                                             & ongoing             & global      & n/a             & NPO         & regulatory, standards                                                    \\
    NEMA Medical Imaging Technology Association (MITA)~\cite{MITAWeb}                                 & ongoing             & USA, global & \textgreater 20 & IF          & regulatory, standards, advocacy                                          \\
    PathLAKE~\cite{PathLAKEWeb}                                                                       & 2019– ongoing       & UK          & \textgreater 20 & PF, IF      & hub, repository, digitalization                                          \\
    PIcc (Pathology Innovation Collaborative Community—Alliance for Digital Pathology)~\cite{PIccWeb} & 2019– ongoing       & USA, global & \textgreater 20 & IF, PF      & regulatory, standards                                                    \\
    \bottomrule
  \end{tabular}
\end{table*}

\subsection{Outlook}

Despite the technical advances and guidelines developed over the past
years, three major hurdles remain for the widespread adoption of AI in
routine pathology. Obtaining regulatory approval remains a
resource-consuming process for AI vendors. This will also be important
to encourage updating products with the newest technologies, which is
particularly relevant in the AI field. The process can be simplified
through better guidance but will continue to require some effort.
Digitization of pathology laboratories is a prerequisite for applying AI
and remains a necessary major up-front investment in equipment and
workflow adaptation by the lab. Enhancing existing guidelines with more
experience and evidence can help to simplify the implementation to some
extent. Reimbursement for using digital or AI-assisted pathology to
cover additional costs, which could exceed what can be saved by more
efficient workflows, will typically need to be decided nationally.
Evidence of value and lobbying work can help make the use of AI
financially viable.

Several standardization and regulatory issues require further
development and sustainable engagement beyond the end of the publicly
funded project. It is for this purpose that we have established the
non-profit EMPAIA International Association. Open to all stakeholders in
the digital pathology ecosystem, this association will remain
vendor-independent. Its funding will come from membership fees,
participation in publicly funded research projects, and the provision of
services to its members, such as scanner benchmarking, provision of
validation data, and interface implementation support. The association
will focus on solving problems that require the cooperation of different
stakeholders and neutrality towards all stakeholders, a prime example
being the successful implementation of open interfaces for pathology-AI
solutions. Therefore, a key goal of the association will be to further
develop the EMPAIA Interfaces to meet new or expanded requirements of AI
solutions and pathology software systems.

The high number of participants, i.e., pathologists, industry,
researchers, legal and quality assurance associations, etc., in the
EMPAIA Academy events, has demonstrated the demand for continuing
education on AI in pathology. The Academy format, open to all interested
parties, will be continued. In addition, the association will offer open
online training programs for using and developing AI solutions with
contributions from leading AI researchers and companies.

EMPAIA International works closely with a growing number of
international reference centers and the stakeholder ecosystem to advance
the use of AI and ensure the practicality of future developments, in
particular by supporting the commercialization of AI prototypes and the
validation of AI solutions. This way, the association will facilitate
the development, marketing, and establishment of AI solutions in
pathology.

\section*{Acknowledgments}

We thank the following investigators for their numerous contributions to
the results presented in this paper: Isil Dogan-O, Stefan Manthey, Arend
Müller, Karl Roth, Katharina Veitengruber (all at Charité --
Universitätsmedizin Berlin); Marcus Barann, Horst Karl Hahn, Henning
Höfener, Johannes Lotz, Daniel Romberg, Daniela Schacherer, Ruben Stein,
Nick Weiss (all at Fraunhofer MEVIS); Andreas Holzinger, Michaela Kargl,
Heimo Müller, Peter Regitnig (all at Medical University of Graz); Thomas
Pilz (at QuIP GmbH); Sahin Albayrak, Theodore Evans, Aray Karjauv, Eric
Kolibacz, Carl Orge Retzlaff, Weijia Shao (all at Technische Universität
Berlin, DAI-Labor); Maren Riemann, and Katharina Wagner (both at
vitasystems GmbH).

Funding: This work was supported by the German Federal Ministry for
Economic Affairs and Climate Action via the EMPAIA project (grant
numbers 01MK20002A, 01MK20002B, 01MK20002C, 01MK20002E, 01MK20002F) and
from the Austrian Research Promotion Agency under grant agreement No.
879881 (EMPAIA). CG has received additional funding via the European
Union’s Horizon 2020 research and innovation programme under grant
agreement No. 101079183 (BioMedAI). TB has received additional funding
from the German Federal Ministry of Research and Education in the
projects PROSurvival (01KD2213D) and Inter-Agent (16DHBKI048). PB has
received additional funding from the European Research Council
(Consolidator Grant No 101001791), the German Federal Ministry of
Education and Research (STOP-FSGS-01GM2202C) and the Innovation Fund of
the Federal Joint Committee (Transplant.KI, No. 01VSF21048). MP has
received funding via the European Union's Horizon 2020 research and
innovation programme under grant agreement No. 101079183 (BioMedAI). TK
has received additional funding from the German Federal Ministry of
Labour and Social Affairs (Go-KI project DKI.00.00032.21).

\section*{Conflicts of Interest}

SS and FZ are employed by an AI vendor of which one app was included in
the evaluation. All other authors declare that they have no conflict of
interest.

\section*{Author Contributions}

Conceptualization: NZ, LOS, TRK, AH, PH; Methodology: NZ, LOS, TK, KW,
TB, PB, RC, MF, CJ, TRK, BL, NCP, KS, FZ, MP, CG, AH, PH; Software: TB,
CJ ,MF, BL, KS, TK; Investigation: LOS, SS, FZ; Writing – Original
Draft: NZ, LOS, TK, KW, TB, RC, MF, CJ, TRK, BL, NCP, KS, MP, CG, AH,
PH; Writing – Review \& Editing: NZ, LOS, TK, KW, TB, PB, RC, MF, CJ,
TRK, BL, NCP, SS, KS, FZ, MP, MT, CG, AH, PH; Visualization: LOS, CJ,
MP, CG; Project administration: PH, NZ, AH, CG, KW, MT; Funding
acquisition: PH, NZ, TRK, AH, CG, KW. All authors read and approved the
final version of the paper.

\bibliography{ms}

\end{document}